\documentclass[journal]{IEEEtran}

\ifCLASSINFOpdf
\else
\fi

\usepackage[absolute]{textpos}
\usepackage[margin=0.52in,top=1in]{geometry}
\usepackage{algorithm}
\usepackage{algpseudocode}
\usepackage{amsmath, amssymb, bm, cite, epsfig}
\usepackage{epstopdf}
\usepackage{graphicx}
\usepackage{array}
\usepackage{multirow}
\renewcommand{\arraystretch}{1.5}
\usepackage{amsfonts}
\usepackage{tabularx} 
\usepackage{tabu}
\usepackage{pbox}
\usepackage{makecell}

\usepackage{booktabs}
\usepackage{ctable}
\usepackage{xcolor,colortbl}
\definecolor{Gray}{gray}{0.85}
\definecolor{LightCyan}{rgb}{0.8,1,1}

\def\beq{\begin{equation}}
\def\eeq{\end{equation}}
\def\beqa{\begin{eqnarray}}
\def\eeqa{\end{eqnarray}}
\def\beqan{\begin{eqnarray*}}
\def\eeqan{\end{eqnarray*}}

\def\PL{\mathrm{PL}}
\def\dB{\mathrm{dB}}

\def\GHz{\mathrm{GHz}}

\def\tm1{t\! - \! 1}
\def\tp1{t\! + \! 1}
\def\log{\mathrm{log}}

\def\PL{\mathrm{PL}}
\def\dB{\mathrm{dB}}

\def\FSPL{\mathrm{FSPL}}
\def\log{\mathrm{log}}
\def\ABG{\mathrm{ABG}}
\def\CI{\mathrm{CI}}

\def\Pr{\mathrm{Pr}}

\def\m{\mathrm{m}}

\def\Pr{\mathrm{Pr}}
\def\Pt{\mathrm{Pt}}
\def\Gr{\mathrm{Gr}}
\def\Gt{\mathrm{Gt}}

\begin{document}
	\begin{textblock}{18.8}(1.5,0.5)
		\centering
		\noindent T. S. Rappaport, S. Sun, and M. Shafi, "Investigation and comparison of 3GPP and NYUSIM channel models for 5G wireless communications," in \textit{2017 IEEE 86th Vehicular Technology Conference (VTC Fall)}, Toronto, Canada, Sep. 2017.
	\end{textblock}
\pagenumbering{gobble}
\title{Investigation and Comparison of 3GPP and NYUSIM Channel Models for 5G Wireless Communications}

\author{\IEEEauthorblockN{Theodore S. Rappaport$^*$, Shu Sun$^*$, and Mansoor Shafi$^\dagger$}\\
	\IEEEauthorblockA{$^*$NYU WIRELESS and NYU Tandon School of Engineering, New York University, Brooklyn, NY, USA\\
		$^\dagger$Spark New Zealand, Wellington, New Zealand\\
		\{tsr, ss7152\}@nyu.edu, Mansoor.Shafi@spark.co.nz}
	
	\thanks{Sponsorship for this work was provided by the NYU WIRELESS Industrial Affiliates program and NSF research grants 1320472, 1302336, and 1555332.}
}
\maketitle
\begin{abstract}
Channel models describe how wireless channel parameters behave in a given scenario, and help evaluate link- and system-level performance. A proper channel model should be able to faithfully reproduce the channel parameters obtained in field measurements and accurately predict the spatial and temporal channel impulse response along with large-scale fading. This paper compares two popular channel models for next generation wireless communications: the 3rd Generation Partnership Project (3GPP) TR 38.900 Release 14 channel model and the statistical spatial channel model NYUSIM developed by New York University (NYU). The two channel models employ different modeling approaches in many aspects, such as the line-of-sight probability, path loss, and clustering methodology. Simulations are performed using the two channel models to analyze the channel eigenvalue distribution and spectral efficiency leveraging the analog/digital hybrid beamforming methods found in the literature. Simulation results show that the 3GPP model produces different eigenvalue and spectral efficiency distributions for mmWave bands, as compared to the outcome from NYUSIM that is based on massive amounts of real-world measured data in New York City. This work shows NYUSIM is more accurate for realistic simulations than 3GPP in urban environments. 
\end{abstract}

\IEEEpeerreviewmaketitle
\section{Introduction}
Fifth-generation (5G) wireless communications will soon be a reality~\cite{Rap13:Access}, and to properly design and deploy 5G wireless systems, channel models based on fundamental physics and extensive measurements at the corresponding frequency bands are needed to analyze future air interfaces and signaling protocols~\cite{Rap_5GTech}. There exist some channel models that have been developed by different groups, such as the 3rd Generation Partnership Project (3GPP), WINNER II, International Telecommunications Union Radiocommunication Sector (ITU-R), METIS (Mobile and wireless communications Enablers for the Twenty-twenty Information Society), COST 2100, MiWEBA (MIllimetre-Wave Evolution for Backhaul and Access) channel models, the statistical spatial channel model NYUSIM developed by New York University (NYU), and others~\cite{3GPP_Dec,Ertel}. 


Among the existing channel models, there are two main types of channel models now being considered by researchers and the industry for 5G wireless: one is the channel model inherited from the model for sub-6 GHz communication systems with modifications to accommodate the spectrum above 6 GHz up to 100 GHz, such as the 3GPP and ITU channel models; the other is the model established based also on extensive propagation measurements at frequencies from 0.5 to 100 GHz, such as NYUSIM developed based on millimeter-wave (mmWave) field measurements~\cite{Rap13:Access,Rap15:TCOM,Samimi15:MTT,Samimi16_Local,Sun17_NYUSIM,Sun16:TVT,Tho16,Mac15_Indoor,Mac17_RMa}. In this paper, multiple-input multiple-output (MIMO) channel eigenvalues are determined from the channel matrices generated from the 3GPP TR 38.900 Release 14~\cite{3GPP_Dec} and NYUSIM~\cite{Samimi15:MTT,Sun17_NYUSIM} models, and spectral efficiencies are studied adopting digital and analog/digital hybrid beamforming methodologies for multi-user (MU) mmWave MIMO systems~\cite{Alk15,Spencer04,Sun17_ICCW}.

\section{Large-Scale Parameter Comparison}
\subsection{LOS Probability Model}
\subsubsection{LOS Probability Model in the 3GPP Channel Model}
The LOS probability models for various scenarios in 3GPP are provided in Table 7.4.2-1 in~\cite{3GPP_Dec}. The LOS probability model is a function of the transmitter-receiver (T-R) separation distance, and sometimes a function of the TX and RX heights. It is inherited from the previous LOS probability model derived for sub-6 GHz bands by 3GPP.

\subsubsection{LOS Probability Model in the NYUSIM Channel Model}
The NYUSIM LOS probability model has a similar form to the 3GPP one, but with the entire formula (i.e., the second equation in Table 7.4.2-1 in~\cite{3GPP_Dec}) squared and the parameter values updated based on statistical modeling from a high resolution ray-tracing approach now described. For a given TX location in Manhattan, a circle was discretized in 100 evenly-spaced points on the circumference around the TX and overlaid on an ariel building map. For each position along the circle external to a building or obstruction, ray-tracing was used to draw a line from the RX to the TX. If that line to the TX penetrated through at least one building, the corresponding initial position at radius $R$ on the circle was denoted as an NLOS position, otherwise it was denoted as a LOS position~\cite{Samimi15_Pro}. This was repeated for all positions along the circle circumference, and the ratio of the number of LOS positions to the total number of positions along the circle provided the LOS probability. This was performed over radii ranging from 10 m to 200 m, in increments of 1 m~\cite{Samimi15_Pro}, and for four TX locations.

\begin{figure}
	\centering
	\includegraphics[width=2.8in]{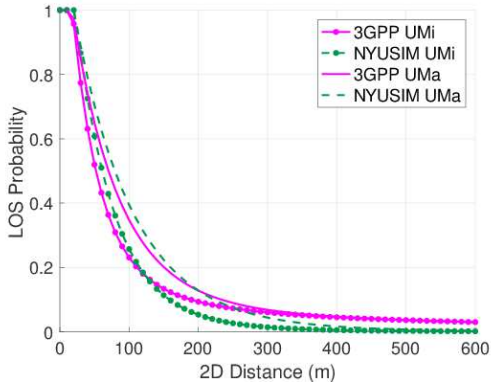}
	\caption{Comparison of LOS probability models in the 3GPP and NYUSIM channel models in UMi and UMa scenarios for a UE height of 1.5 m.}
	\label{fig:LOSProb}
\end{figure}

Fig.~\ref{fig:LOSProb} illustrates the LOS probability models in 3GPP and NYUSIM in UMi and UMa scenarios for a UE height of 1.5 m. As shown by Fig.~\ref{fig:LOSProb}, the 3GPP LOS probability model has a non-zero tail at large distances (several hundred meters), which is not likely to be true in urban environments where numerous tall buildings exist. On the other hand, for T-R separation distances smaller than 160 m, NYUSIM predicts a larger LOS probability compared to 3GPP for both UMi street canyon and UMa scenarios. The difference in the LOS probability impacts spectral efficiency, since LOS facilitates stronger mmWave propagation (i.e., larger SNR) compared to the NLOS condition due to more severe diffraction loss at mmWave than at sub-6 GHz~\cite{Deng16_Diff}. 

\subsection{Large-Scale Path Loss Model}
For a communication link with TX power $\Pt$, TX antenna gain $\Gt$, and RX antenna gain $\Gr$, the received power $\Pr$ [dBm] = $\Pt$ [dBm] + $\Gt$ [dB] + $\Gr$ [dB] - $\PL$ [dB]~\cite{Rap15}, where $\PL$ denotes the large-scale path loss. Two types of large-scale path loss models have emerged in 3GPP: a single-parameter close-in free space reference distance (CI) model and a three-parameter alpha-beta-gamma (ABG) model. Both CI and ABG models are multi-frequency models~\cite{Sun16:TVT}. The CI model, which accounts for the frequency dependence of path loss using a 1 m close-in free space reference distance based on Friis' law~\cite{Friis46a,Andersen95}, can be expressed by the following formula~\cite{Sun16:TVT}:
\begin{equation}\label{CI1}
\begin{split}
&\PL^{\CI}(f_c,d_{3D})[\dB]\\
=&\FSPL(f_c, 1~\m)[\dB]+10n\log_{10}\left(d_{3D}\right)+\chi_{\sigma}^{\CI} \\
=&20\log_{10}\left(\frac{4\pi f_c\times 10^9}{c}\right)+10n\log_{10}\left(d_{3D}\right)+\chi_{\sigma}^{\CI} \\
=&32.4+10n\log_{10}\left(d_{3D}\right)+20\log_{10}(f_c)+\chi_{\sigma}^{\CI}\\
&\text{ where } d_{3D}\geq 1~\m
\end{split}
\end{equation}

\noindent where $\PL^{\CI}(f_c,d_{3D})$ denotes the path loss in dB over frequency and distance, $d_{3D}$ is the 3D T-R separation distance, $\FSPL(f_c, 1~\m)$ denotes the free space path loss in dB at a 1 m 3D T-R separation distance at the carrier frequency $f_c$ in GHz, $n$ denotes the path loss exponent (PLE), $c$ is the speed of light, and $\chi_{\sigma}^{\CI}$ is a zero-mean Gaussian random variable with a standard deviation $\sigma$ in dB~\cite{Sun16:TVT,Andersen95}. Note $\FSPL(1~\GHz, 1~\m)$ is 32.4 dB. The CI model requires just one parameter, i.e., the PLE $n$, to determine the average path loss over distance and frequency, where values for PLEs used by NYUSIM are given in~\cite{Sun17_NYUSIM,Sun16:TVT}. A distinguishing feature of~\eqref{CI1} is that not only does it offer a uniform standard for modeling path loss, but that 10$n$ describes path loss in dB in terms of decades of distances beginning at 1 m, making it very easy to calculate path loss over distance in one's mind. The CI model inherently incorporates frequency dependence of path loss by the FSPL term. 

In addition to the CI path loss model, the CIF (the CI model with a frequency-dependent PLE) and CIH (the CI model with a height-dependent PLE) models, which follow as more general forms of the CI model, are also suitable for multi-frequency path loss modeling, and have been employed in recent work on modeling indoor~\cite{Mac15_Indoor} and rural macrocell (RMa) path loss~\cite{Mac17_RMa}. The general forms of the CIF and CIH path loss models can be found in~\cite{Sun16:TVT,Xing16,Mac15_Indoor,Mac17_RMa}.

The ABG model is expressed as~\cite{Sun16:TVT}:
\begin{equation}\label{ABG1}
\begin{split}
&\PL^{\ABG}(f_c,d_{3D})[\dB]\\
=&10\alpha \log_{10}\left(d_{3D}\right)+\beta+10\gamma \log_{10}\left(f_c\right)+\chi_{\sigma}^{\ABG} \text{,}
\\ &\text{ where } d_{3D}\geq\textrm{ 1 m}
\end{split}
\end{equation}

\noindent where $\PL^{\ABG}(f_c,d_{3D})$ denotes the path loss in dB over frequency and distance, $\alpha$ and $\gamma$ are coefficients embodying the dependence of path loss on distance and frequency, respectively, $\beta$ is an optimized offset value for path loss in dB, $f_c$ is the carrier frequency in GHz, and $\chi_{\sigma}^{\ABG}$ is a zero-mean Gaussian random variable with a standard deviation $\sigma$ in dB describing shadow fading (SF) over large-scale distances. Note that $\gamma$ will be 2 if there is no frequency variation beyond the first meter over all frequencies~\cite{Sun16:TVT}. The ABG model has three model parameters for determining mean path loss over distance and frequency, apart from the shadow fading term, yet offers little to no greater accuracy than~\eqref{CI1}~\cite{Sun16:TVT}.

The path loss models discussed above are omnidirectional path loss models, not directional models. In general, directional path loss models cannot be obtained by simply adding directional antenna gains into the omnidirectional path loss model since not all paths/directions contribute to the directional path loss due to spatial filtering of directional antennas (see Page 3040 of~\cite{Rap15:TCOM}, and~\cite{Sun15_Syn,Mac14_Omni}). However, in a rural scenario with only one cluster/spatial lobe due to lack of scattering objects, directional path loss models are expected to have nearly identical PLEs (with antenna gains represented explicitly outside of the path loss slope with distance, see Eq. (3.9) in~\cite{Rap15}), as compared to omnidirectional path loss models, due to lack of angular diversity in the RMa channel (because there is only one main spatial lobe)~\cite{Mac17_RMa}.

\subsubsection{Large-Scale Path Loss Model in the 3GPP Channel Model~\cite{3GPP_Dec}}
In the UMi street canyon LOS scenario, the CI path loss model is utilized for $d_{3D}$ smaller than the breakpoint distance $d'_{BP}$. After the breakpoint distance, a new term involving the BS and UE heights is added to the CI model, where the BS height is set to 10 m, and the UE height ranges from 1.5 m to 22.5 m. In the UMi street canyon NLOS scenario, the ABG path loss model is adopted with a term accounting for the UE height added to it, while the CI model is listed as an optional path loss model. Similar situations exist in the UMa scenario, except that the BS height is changed to 25 m. 


\subsubsection{Large-Scale Path Loss Model in NYUSIM}
In both UMi and UMa scenarios, the CI model is employed in NYUSIM. The PLE is 2 and 3.2 for UMi LOS and NLOS scenarios, respectively, and 2 and 2.9 for UMa LOS and NLOS, respectively~\cite{Sun16:TVT,Sun17_NYUSIM} (these PLE values can be changed in NYUSIM by the user).

The ABG path loss model adopted in 3GPP is based on legacy approaches that use a regression fit to path loss versus distance for various bands and this is used to derive the intercept and the slope; whilst this may be suitable for evaluation of 5G technologies, it can lead to accuracy problems (see~\cite{Sun16:TVT}). In essence, the 3GPP path loss model has more parameters but generates less accurate results for the long term. The 1-m CI model used in NYUSIM has the same mathematical form with the existing ABG model but has fewer parameters and offers much easier analysis, intuitive appeal, better model parameter stability, and better accuracy over a wide range of microwave and mmWave spectra, scenarios, and distances~\cite{Sun16:TVT}. Different path loss models in 3GPP and NYUSIM impact cell ranges, and the cell range difference between the two channel models varies depending upon the underlying scenario. 

\vspace{-5.4pt}

\subsection{Cluster Definition}
\subsubsection{Clustering Definition in the 3GPP Channel Model}
In the 3GPP model, clusters are characterized by a \textit{joint} delay-angle probability density function, such that a group of traveling multipaths must depart and arrive from a unique angle of departure (AoD)-angle of arrival (AoA) combination centered around a mean propagation delay~\cite{Samimi15:MTT,3GPP_Dec}. High-resolution parameter extraction algorithms, e.g., SAGE and KPowerMeans algorithms~\cite{Fleury99,Czink06} that have high computational complexity, are often employed to obtain cluster characteristics. 

\subsubsection{Clustering Definition in NYUSIM}
NYUSIM uses \textit{time cluster} (TC) and \textit{spatial lobe} (SL) concepts to describe multipath behavior in omnidirectional channel impulse responses (CIRs). TCs are composed of multipath components traveling close in time, and arriving from potentially different directions in a short propagation time window. SLs denote primary directions of departure (or arrival) where energy arrives over several hundred nanoseconds~\cite{Samimi15:MTT}. Per the definitions given above, a TC contains multipath components traveling close in time, but may arrive from different SL angular directions, such that the temporal and spatial statistics are decoupled and can be recovered separately. Similarly, an SL may contain many multipath components arriving (or departing) in a space (angular cluster) but with different time delays. This distinguishing feature is obtained from real-world propagation measurements~\cite{Rap13:Access,Rap15:TCOM} which have shown that multipath components belonging to the same TC can arrive at distinct spatial pointing angles and that energy arriving or departing in a particular pointing direction can span hundreds or thousands of nanoseconds in propagation delay, detectable due to high-gain rotatable directional antennas. The TCSL clustering scheme is physically based, for instance,  it utilizes a fixed inter-cluster void interval to represent the minimum propagation time between possible obstructions causing reflection, scattering, or diffraction, and is derived from field observations based on about 1 Tb of measured data over many years, and can be used to extract TC and SL statistics for any measurement or ray-tracing data sets~\cite{Samimi15:MTT}.

Table~\ref{tbl:ClusterUMi} summarizes the number of clusters and number of rays per cluster in 3GPP, as well as the TCSL statistics in NYUSIM\footnote{The values of parameters given in Table~\ref{tbl:ClusterUMi} are based on 3GPP TR 38.900~\cite{3GPP_Dec}. The 3GPP has recently proposed an alternative channel model in TR 38.901 for frequencies from 0.5 to 100 GHz, which has slightly different path loss models for some scenarios and several other minor differences not related to this paper, but has the same parameter values as in Table~\ref{tbl:ClusterUMi}. The use of the TR 38.901 does not change the results in this paper. TR 38.901 arbitrarily changed the UMa LOS path loss model and eliminated the InH shopping mall model of TR 38.900, reverting back to older sub-6 GHz TR 36.873 without any empirical evidence or explanation.}. It is noteworthy from Table~\ref{tbl:ClusterUMi} that the number of clusters and the number of rays per cluster in 3GPP have fixed values for a given scenario, whereas the number of TCs, the number of subpaths per TC, and the number of SLs (both departure and arrival) in NYUSIM do not hold particular values but follow certain distributions and can vary in each channel realization as shown by extensive field measurements~\cite{Rap13:Access,Rap15:TCOM,Samimi15:MTT,Mac15_Indoor}. More importantly, the 3GPP model has unrealistically large numbers of clusters (e.g., 12 and 19 clusters for the LOS and NLOS environments in the UMi street canyon scenario, respectively), which are not borne out by measurements reported so far and over-predict the diversity of mmWave channels. In contrast, NYUSIM yields only up to 6 TCs and 5 SLs, as these statistics were borne out by extensive measurements in New York City over many years as presented in~\cite{Rap13:Access,Rap15:TCOM,Samimi15:MTT,Rap_5GTech}. 

\begin{table*}
	\renewcommand{\arraystretch}{1.4}
	\caption{Key clustering parameters in the UMi street canyon scenario for frequencies above 6 GHz in the 3GPP and the NYUSIM channel models~\cite{3GPP_Dec,Samimi15:MTT,Sun17_NYUSIM}.}~\label{tbl:ClusterUMi}
	\fontsize{8}{6.5}\selectfont
	\scriptsize
	\begin{center}
		\begin{tabular}{|>{\centering\arraybackslash}m{1.7cm}|>{\centering\arraybackslash}m{4.6cm}|>{\centering\arraybackslash}m{1.8cm}|>{\centering\arraybackslash}m{1.8cm}|>{\centering\arraybackslash}m{1.8cm}|>{\centering\arraybackslash}m{0.6cm}|}\hline
			& \textbf{Parameter Name and Reference Source} & \textbf{LOS} & \textbf{NLOS}  \\ \Xcline{1-4}{1.4pt}
			\multirow{2}{*}{\textbf{3GPP}} & \textbf{Number of clusters}~\cite{3GPP_Dec} & 12 & 19            \\ \cline{2-4}
			& \textbf{Number of rays per cluster}~\cite{3GPP_Dec} & 20 & 20 \\ \Xcline{1-4}{1.4pt}
			\multirow{4}{*}{\textbf{NYUSIM}} & \textbf{Number of time clusters}~\cite{Samimi15:MTT} & \multicolumn{2}{c|}{Discrete Uniform [1, 6]}           \\ \cline{2-4}
			& \textbf{Number of subpaths per time cluster}~\cite{Samimi15:MTT} & \multicolumn{2}{c|}{Discrete Uniform [1, 30]} \\ \cline{2-4}
			& \textbf{Number of spatial lobes (departure)}~\cite{Samimi15:MTT} & Poisson(1.9) & Poisson(1.5) \\ \cline{2-4}
			& \textbf{Number of spatial lobes (arrival)}~\cite{Samimi15:MTT} & Poisson(1.8) & Poisson(2.1) \\ \cline{1-4}
		\end{tabular}
	\end{center}
\end{table*}

There are also discrepancies in other large-scale parameters between 3GPP and NYUSIM, but they are not detailed here due to space limit. Please refer to~\cite{3GPP_Dec} and~\cite{Samimi15:MTT,Samimi16:EuCAP} for detailed information. 

\begin{figure*}[t]
	\centering
	\includegraphics[width=6in]{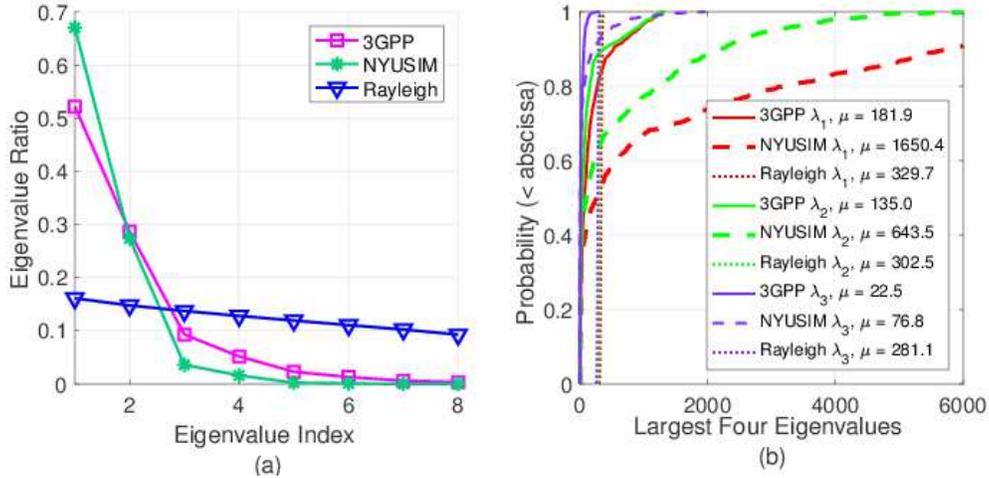}
	\caption{Comparison of (a) channel eigenvalue ratios of the largest eight eigenvalues, and (b) CDFs of the largest three eigenvalues in the 3GPP~\cite{3GPP_Dec} and NYUSIM~\cite{Sun17_NYUSIM,Samimi15:MTT} channel models in the UMi street canyon scenario for one user equipment (UE) in a single cell, as well as a Rayleigh fading channel. The eigenvalue ratio is obtained by dividing the eigenvalue by the sum of all the eigenvalues in linear scale. $\lambda$ denotes the eigenvalue, and $\mu$ denotes the mean of the eigenvalues.}
	\label{fig:EV}
\end{figure*}

\section{Small-Scale Parameter Comparison}
The 3GPP and NYUSIM models also have different small-scale parameters (SSPs) (see~\cite{3GPP_Dec} and~\cite{Samimi15:MTT,Samimi16:EuCAP} for more detailed information). The SSPs impact the channel impulse response and therefore the eigenvalues which in turn affect spectrum efficiency. In this section, the 3GPP directional antenna element radiation pattern in Table 7.3-1 in~\cite{3GPP_Dec} is applied to both 3GPP and NYUSIM to generate the channel matrix \textbf{H} according to the form in Eq. (7.5-28) in~\cite{3GPP_Dec} to study channel eigenvalues and spectrum efficiency. 

 Let us assume a single-cell MU-MIMO system operating at 28 GHz with an RF bandwidth of 100 MHz in the UMi street canyon scenario. Even though 5G systems will have large bandwidths (up to 1 GHz), this bandwidth is likely to be aggregated over component carriers each 100 MHz wide or more. Bandwidths larger than 100 MHz will result in SNR decrease at the cell edge, which could be made up with higher gain directional antennas. The BS is equipped with a uniform rectangular array (URA) with 256 cross-polarized antenna elements (where 128 elements are $+45^\circ$ slanted, and the other 128 are $-45^\circ$ slanted), and each UE has 8 antenna elements constituting a URA with cross-polarized omnidirectional elements (where 4 elements are $+45^\circ$ slanted, and the other 4 are $-45^\circ$ slanted). Cross-polarized antenna elements are considered herein since they can effectively reduce the physical size while making use of different polarization components. Each BS antenna element has a radiation pattern as specified in Table 7.3-1 of~\cite{3GPP_Dec} with a maximum gain of 10 dB, and each UE antenna element possesses an omnidirectional pattern. In the simulations, it is assumed that 95\% of the area in the cell have an SNR larger than or equal to 5 dB, and the upper bound of the T-R separation distance is calculated based on this assumption by manipulating the SF in~\eqref{CI1} and~\eqref{ABG1}, such that the cell radius corresponds to a SF that splits the area under the SF standard deviation probability density function (PDF) into two parts which occupy 95\% and 5\% of the total area under the PDF, respectively. Users are randomly dropped with a T-R separation distance between the lower bound (10 m) and the upper bound of the distance based on the SNR threshold described above. All the simulation results are obtained from 1000 random channel realizations (i.e., 1000 random user drops).

Fig.~\ref{fig:EV} illustrates the channel eigenvalue ratios of the largest eight eigenvalues and cumulative distribution functions (CDFs) of the largest three eigenvalues for one user using the 3GPP and NYUSIM channel models in the UMi street canyon scenario, superimposed with the results for a Rayleigh fading channel. The eigenvalue ratio is obtained by dividing the eigenvalue by the sum of all the eigenvalues in linear scale. As seen from Fig.~\ref{fig:EV}(a), the eigenvalues for the Rayleigh fading channel are relatively close to each other, while the ordered eigenvalues for 3GPP and NYUSIM decrease quickly, especially for NYUSIM. For example, the largest eigenvalue is about 1.7 and 153.5 times the smallest one in Rayleigh and 3GPP channels, respectively, whereas the largest eigenvalue is more than 9564 times the smallest one in NYUSIM. With regards to the absolute value and distribution of the eigenvalues, Fig.~\ref{fig:EV}(b) shows that the largest eigenvalue yielded by NYUSIM is about an order of magnitude higher on average than that generated by 3GPP, and is about five times the largest eigenvalue for a Rayleigh fading channel. For the second largest eigenvalue, similar trends hold but with smaller discrepancies among the three types of channels. As to the third largest eigenvalue, the value is higher in NYUSIM than in 3GPP, but both of them are substantially smaller than what the Rayleigh fading channel produces. This reveals that measured mmWave channels generated by NYUSIM have only a few but strong dominant eigenmodes, whereas the 3GPP model yields more eigenmodes with weaker powers in dominant eigenmodes when compared to NYUSIM~\cite{Rap_5GTech}.

\begin{figure}
	\centering
	\includegraphics[width=3in]{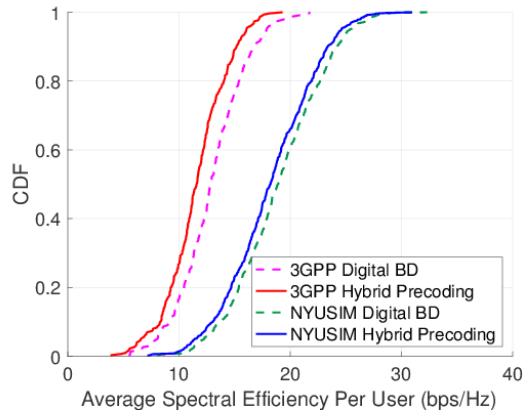}
	\caption{CDFs of the spectral efficiency per user (averaged over three users) for a single-cell three-user MIMO system for both the 3GPP and NYUSIM channel models, using the hybrid precoding and analog combining method proposed in~\cite{Alk15} and the digital block diagonalization (BD) approach presented in~\cite{Spencer04} with one data stream and one RF chain per user.}
	\label{fig:SE}
\end{figure}

The CDFs of the spectral efficiency per user for a single-cell three-user MIMO system are depicted in Fig.~\ref{fig:SE} for both 3GPP and NYUSIM, using the hybrid precoding and analog combining method proposed in~\cite{Alk15} and the digital block diagonalization (BD) approach presented in~\cite{Spencer04} with one data stream and one RF chain per user. In the simulations using the hybrid precoding method, both the BS and UE beamforming codebooks are composed of the antenna array response vectors corresponding to the actual AoDs and AoAs generated from the channel model being investigated, and quantization error in the feedback stage is not considered. It is observed from Fig.~\ref{fig:SE} that for both the hybrid precoding and the digital BD methods, NYUSIM yields higher spectral efficiencies than 3GPP, due to the larger dominant eigenvalue generated by NYUSIM. Specifically, at the 50\% point in the CDF, there is a 6 bits-per-second-per-hertz (bps/Hz) gap in the spectral efficiencies between NYUSIM and 3GPP models. Furthermore, the spectral efficiency disparity between the digital BD and the hybrid precoding approaches is smaller in NYUSIM. The most possible reason for this is that the multi-user digital precoding matrix in~\cite{Alk15} is obtained through the zero-forcing (ZF) method, since the channel generated by NYUSIM is sparser compared to 3GPP, thereby the effective channel after RF precoding/combining is more likely to be well-conditioned so as to make ZF capable of achieving near-optimal performance~\cite{Yoo06,Alk15}. 

\section{Conclusion}
Two popular channel models, i.e., the 3GPP~\cite{3GPP_Dec} and NYUSIM~\cite{Sun17_NYUSIM,Samimi15:MTT} models, are investigated and compared in terms of their modeling methodologies and channel evaluation performance. In contrast to the 3GPP model that relied on many legacy sub-6 GHz results, NYUSIM emphasizes more physical basis and builds upon massive amounts of true measured data at mmWave frequencies~\cite{Rap13:Access,Rap15:TCOM,Samimi16_Local,Samimi15:MTT,Samimi15_Pro,Sun17_NYUSIM,Tho16,Rap_5GTech}.

Simulations leveraging the analog/digital hybrid beamforming approach for a MU-MIMO scenario are conducted, to explore practical mmWave systems where numerous antenna elements can be packed into a small form factor. Simulation results show NYUSIM yields stronger and fewer dominant eigen-channels per user than 3GPP, due to the much smaller number of TCs and SLs (sparsity) that has been borne out by field measurements. Moreover, the spectral efficiency per user for a single-cell three-user MIMO system generated by NYUSIM is about 1.5 times that produced by 3GPP considering the median point, because of the larger dominant eigenvalue in the former. This paper shows that NYUSIM offers more realistic simulation results  than 3GPP,  and this has a major impact on capacity and implementation aspects. Future work will investigate multi-stream and multi-cell MU-MIMO performance using the two channel models.


\ifCLASSOPTIONcaptionsoff
  \newpage
\fi

\bibliographystyle{IEEEtran}
\bibliography{bibliography}

\end{document}